\documentstyle[ltwol,psfig]{article}

\def\be{\begin{equation}}
\def\ee{\end{equation}}
\def\bea{\begin{eqnarray}}
\def\eea{\end{eqnarray}}

\begin{document}

\title{COMPARISON OF NEUTRINO AND MUON STRUCTURE FUNCTIONS, SHADOWING 
CORRECTIONS AND CHARGE SYMMETRY VIOLATION}

\author{C. BOROS AND A. W. THOMAS }

\address{Department of Physics and Mathematical Physics,
                and Special Research Center for the
                Subatomic Structure of Matter,
                University of Adelaide,
                Adelaide 5005, Australia}

\author{J. T. LONDERGAN}

\address{ Department of Physics and Nuclear
            Theory Center, Indiana University,
            Bloomington, IN 47404, USA}


\twocolumn[\maketitle\abstracts{ 
Comparison of structure functions measured in neutrino and charged 
lepton deep inelastic scattering can be used to test basic 
properties of parton distribution functions such as 
the validity of charge symmetry. Recent experiments indicate 
a substantial discrepancy between $F_2^\nu$ and $F_2^\mu$ in the 
region of small Bjorken-x. We discuss nuclear corrections and 
strange and anti strange quark effects and show that none of them  
can account for the observed discrepancy.  
These results suggest surprisingly 
large CSV effects in nucleon sea distributions.}]

\section{Introduction}

Comparison of structure functions measured  in neutrino  
and charged lepton deep inelastic scattering       
can be used to  test basic symmetry properties of parton 
distribution functions such as the validity of charge symmetry 
\cite{Miller}. 

Such comparisons are based on the interpretation of these 
structure functions in terms of parton distribution functions. 
In the quark-parton model the structure functions measured in
neutrino, anti neutrino  and charged lepton DIS on an iso-scalar
target, $N_0$, are given by \cite{Lon}  
\begin{eqnarray}
  F_2^{\nu N_0} (x) &=& x[ Q(x)     
     + 2 s(x)  -\delta u(x)-\delta \bar d(x)] \nonumber \\
  F_2^{\bar \nu N_0} (x) &=& x[Q(x) 
     + 2 \bar s(x) -\delta d(x)-\delta \bar u(x)] \nonumber \\
  F_2^{\ell N_0}(x) & =& \frac{5}{18} x
   [ Q(x) + 
\frac{2}{5} (s(x) + \bar s(x))
  \nonumber\\
 & -  & \frac{4 
 (\delta d(x)+\delta \bar d(x))}{5} - \frac{( \delta u(x)+\delta
   \bar u(x))}{5}]
\label{eq2}
\end{eqnarray}
Here, we introduce the notation $Q(x)=u(x)+\bar u(x) 
+d(x) +\bar d(x)$ and express everything in terms of 
parton distributions functions in the proton and  
charge symmetry violating distributions which are defined as  
the differences between the up (down) quark distribution 
in  the proton and the down (up) quark distribution in the neutron 
\begin{eqnarray}
\delta u(x)& =&  u^p(x) -d^n(x) \nonumber\\
\delta d(x)& =&  d^p(x) -u^n(x). 
\end{eqnarray}
If charge symmetry is valid these terms are zero. 

A very useful quantity in the comparison of the structure functions 
is  the ``charge ratio'' 
\begin{eqnarray}
 R_c (x) & \equiv  & \frac{F_2^{\mu N_0}(x)}{\frac{5}{18}
 F_2^{\nu N_0}(x) -x[ s(x) +\bar s(x)]/6}.  
\label{rc}
\end{eqnarray}
A deviation $R_c(x) \ne 1$, at any value of $x$, must arise either
from CSV effects or from $s(x) \ne \bar{s}(x)$.  

Recent experimental measurements allow a precise comparison between
$F^\nu_2(x,Q^2)$ and $F_2^{\mu}(x,Q^2)$.  
In Fig.\ref{fig1} we show the ``charge ratio'' calculated 
by using the CCFR \cite{CCFR}
and NMC \cite{NMC} structure functions for 
$F_2^\nu$ and $F_2^\mu$, respectively.  The structure functions 
are integrated over $Q^2=3.2$ GeV$^2$ in the 
region of overlap  of the two experiments. 
Since the CCFR structure  function has been measured on an iron 
target nuclear  corrections  
(nuclear EMC effect, shadowing and
anti shadowing)  have to be applied to the data. 
In Fig. 1, the open triangles represent the result 
without nuclear corrections.  
The open circles are calculated by  using nuclear 
corrections obtained from {\it charged lepton} DIS.    
While, in the region of intermediate values of
Bjorken $x$ ($0.1 \le x \le 0.4$), the two structure functions are
in very good agreement, in the small
$x$-region ($x < 0.1$), they differ by as much as 10-15$\%$ 
when nuclear corrections are taken into account. 
This discrepancy could be interpreted as evidence for 
charge symmetry violation. 
However, several corrections have to be applied to the data before any
conclusions may be drawn. 
We see especially that the result is very sensitive to nuclear  
corrections.  
The CCFR Collaboration made a careful study of overall normalization,
charm threshold and iso-scalar correction effects \cite{CCFR}. Here, we 
re-examine nuclear corrections for neutrinos and 
discuss $s(x) \ne \bar{s}(x)$ effects before we turn to  
possible  charge symmetry violation.

\section{Nuclear corrections} 

Nuclear  corrections for neutrinos 
are generally calculated
using correction factors from charged lepton reactions
at the same kinematic values. {\it A priori}, there is no reason
that neutrino and charged lepton heavy target corrections should be
identical, especially if such corrections
depend  strongly on the properties of
the exchanged object (photon, W) used to probe the structure of the target.
Since this is the case for  nuclear shadowing
corrections in the small $x_B$ region
for small to moderately large $Q^2$-values
we re-examined shadowing corrections to  neutrino DIS
focusing on the differences between neutrino and charge
lepton scattering and on effects due to the $Q^2$-dependence of
shadowing.  We used a two phase model
which has been successfully applied to the description of shadowing
in charged lepton DIS
\cite{Badelek}.
In generalizing this approach to weak currents,  subtle
differences  between shadowing in neutrino and charged lepton DIS
arise because of the partial conservation of axial currents (PCAC)
and the coupling of  the weak current to both vector and axial vector mesons.
PCAC requires that  that the divergence of the axial current does not vanish
but is proportional to the pion field for $Q^2=0$. This
is Adler's theorem \cite{Adler},
which relates the neutrino cross section to the pion
cross section  on the same target for $Q^2=0$.
Thus, for low $Q^2\approx m_\pi^2$ shadowing in neutrino
scattering is determined by the absorption of  pions on the target.
\begin{figure}
\center
\psfig{figure=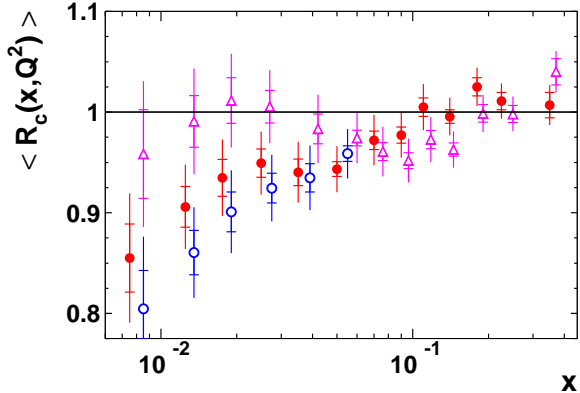,height=2.4in}
\caption{The ``charge ratio'' $R_c$ vs.\ $x$  calculated using
         CCFR \protect\cite{CCFR} data for neutrino and
         NMC \protect\cite{NMC} data for muon
         structure functions. Open triangles: no heavy target
         corrections; open circles: $\nu$ data corrected for heavy
         target effects using corrections from charged lepton scattering;
         solid circles: $\nu$ shadowing corrections calculated in the
         ``two phase'' model. 
          Both statistical and systematic errors are shown. }
\label{fig1}
\end{figure}

For larger $Q^2$-values the contributions of vector and axial vector mesons
become important. The coupling of the weak current
to the vector and axial
vector mesons and that of the electro-magnetic current
to vector mesons are related to each other by the ``Weinberg sum rule''
$f_{\rho^+}^2=f_{a_1}^2=2f_{\rho^0}^2$.
Since the coupling of the vector (axial vector)
mesons to the weak current is  twice as large as  the coupling
to the electro-magnetic current but the structure function is
larger by a factor of $\sim18/5$  in the neutrino
case, we expect that shadowing due to VMD in neutrino reactions is
roughly half of that in charged lepton scattering. 
For large $Q^2$-values, 
shadowing due to Pomeron exchange (which is of leading twist)
becomes dominant, leading to identical
(relative) shadowing in neutrino and charged lepton DIS.

We calculated shadowing corrections \cite{Boros} 
to the CCFR $\nu$ structure function 
in the framework of this ``two phase'' model and
used these corrections in calculating the charge ratio $R_c$ of
Eq.\ \ref{rc}. 
There are also nuclear effects in the Deuteron. However,
because of the low density of the Deuteron,
these are (relatively speaking)
very small and have a negligible effect on the charge ratio.  
We integrated the structure functions
above $Q^2=3.2$ GeV$^2$ in the overlapping kinematic region
of the two experiments and
used a parametrization of the nuclear corrections
in charged lepton DIS to correct the data in the non-shadowing region.
The result is shown as solid circles  in
Fig.\ref{fig1}.  
At small $x$, careful consideration of neutrino shadowing corrections
decreases, but does not resolve, the low-$x$ discrepancy between the
CCFR and NMC data.

\section{Strange and anti strange quark effects}

Since the CCFR-Collaboration uses both neutrino and 
anti neutrino events in the structure function analysis 
the extracted  structure function $F_2^{CCFR}$ is a flux weighted average
between $\nu$ and $\bar{\nu}$ structure functions \cite{CCFR} 
\begin{eqnarray}
  F_2^{CCFR}(x,Q^2)& = &\alpha F_2^\nu (x,Q^2) + (1-\alpha )
F_2^{\bar\nu}(x,Q^2).
\label{s1}
\end{eqnarray}
Here, we define the relative neutrino 
flux as $\alpha = \Phi_\nu /(\Phi_\nu + \Phi_{\bar\nu})$, where
$\Phi_\nu$ and $\Phi_{\bar\nu}$ are the  $\nu$ and $\bar{\nu}$
fluxes, respectively.  
$F_2^{CCFR}$ is equal to $\frac{1}{2}[ F_2^\nu(x,Q^2)+
F_2^{\bar\nu}(x,Q^2)]$ if $\alpha=\frac{1}{2}$ or if
the two structure functions are equal, i.e. charge symmetry is valid 
and  $s(x) =\bar s(x)$.   
The value of $\alpha$ depends on the energy of the incident 
neutrinos and anti neutrinos in the CCFR-experiment.  
In the relevant kinematic region, at small $x$ which corresponds 
to large incident energies,  it is  $\approx 0.83$  
so to a good approximation $F_2^{CCFR}(x,Q^2)$ can be regarded as a neutrino
structure function. 
\begin{figure}
\psfig{figure=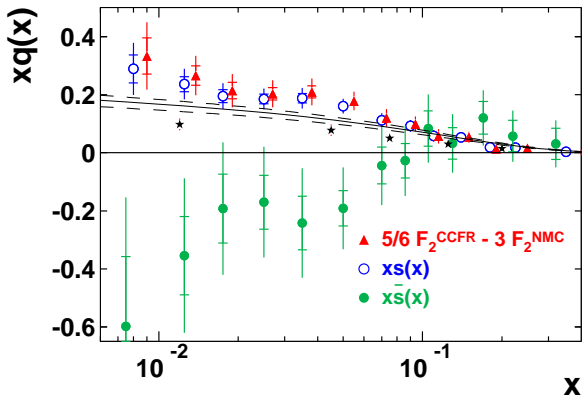,height=2.8in}
\vspace*{-1.cm} 
\caption{$x s(x)$ (open circles) and $x \bar{s}(x)$ (solid circles)
      extracted by combining CCFR and NMC structure functions
      with $s(x)^{\mu\mu}$ extracted from dimuon experiments, as given in
      Eqs.\ \protect\ref{diff} and \protect\ref{s2}.
      Solid triangles: $\frac{5}{6} F_2^{CCFR}-3F_2^{NMC}$. 
      Stars: $x s(x)$ from a LO-analysis \protect\cite{CCFRLO}.  
      Solid line: $x s(x)$ from a NLO-analysis 
       \protect\cite{CCFRNLO}; dashed band indicates $\pm 1\sigma$
      uncertainty.}
\label{fig2}
\end{figure}

The remaining small-$x$ discrepancy in the
charge ratio is either from different strange quark
distributions $s(x) \ne \bar{s}(x)$ 
or from charge symmetry violation. First, we examine
the role played by the strange quark distributions. Assuming charge
symmetry, $s(x)$ and $\bar{s}(x)$ are given by a
linear combination of neutrino and muon structure functions,
\begin{eqnarray}
   \frac{5}{6} F_2^{CCFR}(x,Q^2) && - 3 F_2^{NMC}(x,Q^2)
= \frac{1}{2}\, x \, [s(x) + \bar s(x)]\nonumber \\
&& +\frac{5}{6} (2\alpha -1)\, x \,[s(x)-\bar s(x)].
\label{diff}
\end{eqnarray}
Under the assumption $s(x)=\bar s(x)$, this relation could be
used to extract the strange quark distribution.
However, as is well known, $s(x)$ obtained in this way is inconsistent
with results extracted from independent experiments.

Opposite sign dimuon production in deep inelastic $\nu$ and
$\bar{\nu}$ scattering provides a direct determination of both
$s(x)$ and $\bar{s}(x)$. The CCFR Collaboration extracted $s(x)$
and $\bar{s}(x)$ from a leading order (LO) \cite{CCFRLO} 
next to leading order (NLO) analysis
\cite{CCFRNLO} of their dimuon data. 
While the strange and anti strange
distributions were different in the LO analysis they were  
equal within experimental errors in the NLO analysis. 

To test the hypothesis
that  the low-$x$ discrepancy
in the charge ratio of Eq.\ \ref{rc} could be accounted for by allowing
$s(x) \ne \bar{s}(x)$ we combined the 
dimuon production data, averaged over $\nu$ and
$\bar\nu$ events, with the structure functions from neutrino and
charged lepton scattering (Eq. \ref{diff}).
Defining $\alpha^\prime= N_\nu/(N_\nu+N_{\bar\nu})$, where
$N_\nu =5,030 $, $N_{\bar\nu}=1,060$ ($\alpha^\prime \approx 0.83$)
are respectively the $\nu$ and $\bar\nu$
events from the dimuon production experiment \cite{CCFRNLO},
the flux-weighted experimental distribution $x s(x)^{\mu\mu}$ from
dimuon production is
\begin{eqnarray}
 x s^{\mu\mu}(x) &= &\frac{1}{2}\, x \,[s(x) + \bar s(x)] \nonumber\\ 
&+& \frac{1}{2}
       (2\alpha^\prime -1 )\, x\, [s(x) - \bar s(x)].
\label{s2}
\end{eqnarray}
Eqs. (\ref{diff}) and (\ref{s2})
form a pair of linear equations which can be solved for
$s(x)$ and  $\bar s(x)$. 
We can simultaneously test the compatibility of the various
experiments. 

Compatibility of the two experiments requires that physically
acceptable solutions  for
$\frac{1}{2} x [s(x)+\bar s(x)]$ and  $\frac{1}{2} x [s(x)-\bar s(x)]$,
satisfying both Eq. \ref{diff} and Eq. \ref{s2}, can be found.
Clearly, solutions do not exist if $\frac{5}{3}
(2\alpha -1)=(2\alpha^\prime -1)$. (Note that the left-hand 
sides of Eq. \ref{diff} and Eq. \ref{s2} are different.) 
This gives the critical values $\alpha =\frac{1}{5} (3\alpha^\prime +1)$
or $\alpha \approx 0.7$ for $\alpha^\prime \approx 0.83$
and $\alpha = 0.8$ for $\alpha^\prime =1$. 

In Fig.\ \ref{fig2} we show the results obtained for $x s(x)$ (open
circles) and $x \bar s(x)$ (solid circles) by solving the resulting
linear equations, Eqs.\ \ref{diff} and \ref{s2}, with 
$\alpha\approx\alpha^\prime\approx 0.83$.  
Both the structure functions and dimuon data
have been integrated over $Q^2>3.2$ GeV$^2$ in the overlapping
kinematical regions.
In averaging the dimuon data we used the CTEQ4L parametrization for
$s^{\mu\mu}(x)$ \cite{Lai}. We see that the results are
completely unphysical, since the equations require $\bar{s}(x) < 0$.  
Since $\alpha \approx 0.83$ is closer to the critical value  
for $\alpha^\prime =1$ than for  $\alpha^\prime =0.83$   
the results is even more unphysical if we use only the 
neutrino dimuon data. Similarly, if we decrease $\alpha$ 
$\bar s(x)$ becomes more negative.  

In conclusion, our analysis strongly suggests that
requiring charge symmetry, but allowing $s(x) \ne \bar{s}(x)$,
cannot resolve the discrepancy between $F_2^{CCFR}(x,Q^2)$ and
$F_2^{NMC}(x,Q^2)$.  The experimental results are incompatible,
even if $\bar{s}(x)$ is completely unconstrained.

\section{Charge symmetry violation} 

We have seen that 
neither neutrino shadowing corrections nor allowing $s(x) \ne
\bar{s}(x)$ removes the low-$x$ discrepancy. There remain the two possible
explanations.  Either one of the experimental
structure functions (or $s(x)$) is incorrect
at low $x$, or parton charge symmetry is violated in this
region. Assuming the possibility of parton CSV, we
can combine the dimuon data for $s(x)$, (Eq.\ \ref{s2}), with
Eq.\ \ref{diff} to obtain 
\begin{eqnarray}
& &   \frac{5}{6} F_2^{CCFR}(x,Q^2) - 3 F_2^{NMC}(x,Q^2)
 -x s^{\mu\mu}(x) \nonumber \\ 
&& \approx \frac{1}{2} [\frac{5}{3}(2\alpha -1) -(2\alpha^\prime -1)] 
x[s(x) -\bar{s}(x)] \nonumber \\ 
& &  \,\,\,\,\,\,\,\,\,\,\, +\frac{1}{2} \, x \, [\delta \bar d(x) -\delta
\bar u(x)].
\label{csv1}
\end{eqnarray} 
This equation is valid at small $x$, where
sea quark distributions are much larger than valence quarks, so we make
the simplest assumption, namely that
$\delta q_v(x)=\delta q(x) -\delta \bar q(x) \approx 0$.  
\begin{figure}
\psfig{figure=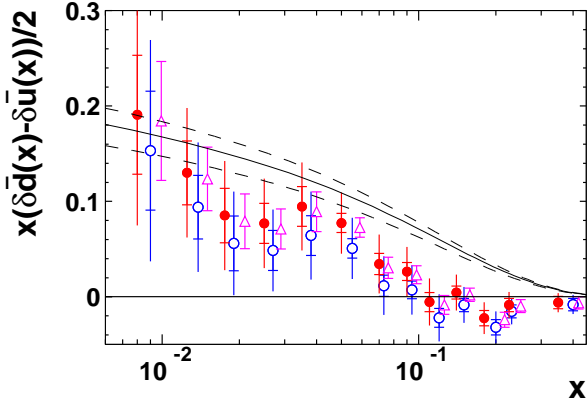,height=2.8in}
\vspace*{-1.cm} 
\caption{Charge symmetry violating distributions extracted
      from the CCFR and NMC structure function data and
      the CCFR dimuon production data under the assumption
      that $s(x)=\bar s(x)$ (solid circles) and
      $\bar s(x)\approx 0$ (open circles) for 
        $\alpha^\prime =0.83$ and $s(x)=\bar s(x)$ (solid circles) and
      $\bar s(x)\approx 0$ (open triangles) for $\alpha^\prime =1$. 
      (For the latter only statistical errors are shown.)  
       $s(x)$ at $Q^2=4$ GeV$^2$ obtained by the  CCFR 
       Collaboration in a NLO analysis 
     \protect\cite{CCFRNLO} is shown for comparison.  }
\label{fig3}
\end{figure}

The left hand side of Eq.\ \ref{csv1} and the 
coefficient in front of $s$-$\bar{s}$ are positive.  Consequently,
the smallest CSV effects will be obtained when
$\bar{s}(x) = 0$. Note that this violates the requirement that 
the net strangeness of the nucleons be zero.  
Thus, setting $\bar{s}(x) = 0$ gives the absolute 
lower limit on charge symmetry violation. 
In Fig. \ref{fig3} we show the
CSV effects needed to satisfy the
experimental values in Eq.\ \ref{csv1}. The open circles
are obtained when we set $\bar{s}(x) = 0$, and the solid circles
result from setting $\bar{s}(x) = s(x)$.  
We also show the results  
we obtain by setting $\alpha^\prime =1$ which corresponds to 
using only a subsample of the di-muon data containing only 
neutrino events. For $\bar{s}(x) = s(x)$, the result is the same. 
However, setting $\bar{s}(x) = 0$  gives us a higher lower limit 
on CSV  than that for $\alpha^\prime$.  The result is shown as 
open triangles in Fig. \ref{fig3}. 
  
\begin{figure}
\psfig{figure=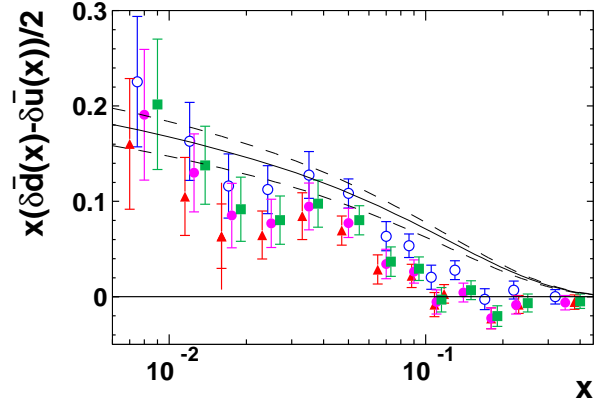,height=2.8in}
\vspace*{-1.cm} 
\caption{Uncertainty due to the parametrization used 
for the dimuon data on charge symmetry violation. 
Open circles: LO CCFR distribution, solid circles: 
CTEQ4L, solid rectangles: CTEQ4D, solid triangles: NLO CCFR distribution.
Here, except for the most ``critical'' point, 
only statistical errors are shown. 
}
\label{fig4}
\end{figure}

The magnitude of the extracted CSV is also 
sensitive to the strange quark distribution used in the analysis. 
In Fig.\ref{fig4} we show the  CSV obtained by using the LO 
(open circles) and NLO (solid triangles)  
CCFR distributions and the CTEQ4L (solid circles) and CTEQ4D 
(solid rectangles) parametrizations. 
The uncertainty due to different parametrizations 
has been partly  taken into account since the calculated errors already 
include the uncertainty of the dimuon measurement and most of the  
parametrizations lie within the experimental 
errors of the dimuon data (except of LO-CCFR $s(x)$).

In conclusion, the CSV effect required 
to account for the low-$x$ NMC-CCFR discrepancy is extraordinarily large.
It is roughly the same size as the strange quark distribution at
small $x$. This CSV term is roughly 25\% of the
light sea quark distributions for $x < 0.1$, and the sign
gives $\bar{d}^p(x) > \bar{u}^n(x)$ and  $\bar{d}^n(x) > \bar{u}^p(x)$.

\section{Influence of CSV on other observables}

Clearly, CSV effects of this magnitude need further experimental
verification.  It is hard to imagine how such large CSV effects are
compatible with the high precision of charge symmetry measured
at low energies.  The level of CSV required is surprising, as it is
at least two orders of magnitude larger than theoretical CSV estimates
\cite{Sather,Ben98}. Theoretical considerations suggest 
that $\delta\bar d(x) \approx -\delta\bar u(x)$ \cite{Ben98}; 
with this sign CSV 
effects also require large flavor symmetry violation.
Since most of the observables are proportional to 
the sum of $\delta\bar d(x)$ and $\delta\bar u(x)$ rather than to their  
difference this large CSV could remain unobserved in many 
experiments.  Here, we briefly discuss the effects of the extracted 
CSV on FSV. 

\begin{figure}
\psfig{figure=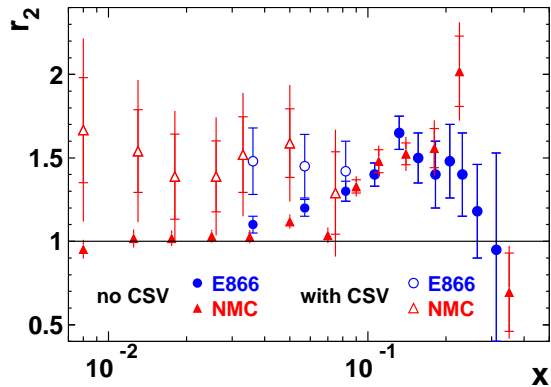,height=2.8in}
\vspace*{-1.cm} 
\caption{The ratio $\bar d/\bar u$
extracted from the Drell-Yan data assuming the validity of
charge symmetry (solid circles).
If CS is violated this ratio corresponds to
\protect$(\bar d-\delta \bar d)/\bar u$. The result obtained by correcting
for CSV is shown as open circles.
The same for the ratio extracted from the difference of the
proton and Deuteron structure functions is shown as solid and
open triangles, with and without CSV, respectively. }
\label{fig5}
\end{figure}

The Drell-Yan experiment measures
the ratio of the dimuon cross sections of proton-Deuteron
and proton-proton scattering. Since CSV is significant in the small
$x$ region, it is a good approximation 
to keep only the  contributions to the Drell-Yan cross sections
which come from the annihilation of
quarks of the projectile and anti quarks of the target.
In this approximation, the ratio $R\equiv \sigma^{pD}/(2\sigma^{pp})$
is given by
\begin{equation}
\frac{\sigma^{pD}}{2\sigma^{pp}}
\approx \frac{1}{2}  \frac{[1+\frac{\bar d_2}{\bar u_2} -
  \frac{\delta\bar d_2}{\bar u_2}] +\frac{R_1}{4}[1+\frac{\bar d_2}{\bar u_2}
-\frac{\delta\bar u_2}{\bar u_2}]}{1+\frac{R_1}{4}\frac{\bar d_2}{\bar u_2}} .
\end{equation}
Here, we introduced the notation  $R_1\equiv \frac{d_1}{u_1}$ and
$q_{1,2}\equiv q(x_{1,2})$ for the quark distributions.
Neglecting $R_1$ for large $x_F$, which corresponds to large $x_1$, we have
\begin{equation}
  R =\frac{\sigma^{pD}}{2\sigma^{pp}}
   \approx  \frac{1}{2} \{ 1 + \frac{(\bar d_2 -
\delta\bar d_2)}{\bar u_2} \} .
\label{r}
\end{equation}
If charge symmetry is violated, the extracted quantity 
\cite{E866} is not
$\bar d_2/\bar u_2$  but
$r_2\equiv (\bar d_2 -\delta\bar d_2)/\bar u_2$.
Since $\delta\bar  d$ is positive the FSV ratio, $\bar d_2/\bar u_2$,
should be enhanced for small $x$ relative to the measured value
$r_2$. The enhancement is in the order of
$25\%$ in the small $x$ region where CSV could be important. 
This is shown in Fig. \ref{fig5}. 

The difference, $\bar d -\bar u$, can also be extracted
from the difference between the proton and neutron structure
functions measured by the the NMC Collaboration \cite{NMC}. 
In this case we have
\begin{equation}
   \frac{1}{2}(u_v-d_v)-\frac{3}{2x}(F_2^p-F_2^n) =
   (\bar d -\bar u ) -\frac{2}{3}(\delta d +\delta \bar d)
  -\frac{1}{6} (\delta u +\delta\bar u).
\end{equation}
We can make the approximations $\delta q\approx \delta\bar q$
and $\delta\bar d \approx -\delta\bar u$,  
and obtain
\begin{equation}
   \frac{1}{2}(u_v-d_v)-\frac{3}{2x}(F_2^p-F_2^n) \approx
   [(\bar d -\delta\bar d)-\bar u ].
\label{nmcdiff}
\end{equation} 
We see that, in a first approximation,    
the quantities extracted from the two experiments are the same even if
both CSV and FSV are present. However, if CSV is present, the
term $\delta\bar d$ has to be subtracted from the measured
quantity to obtain the difference $\bar d -\bar u$.

We inverted Eq. \ref{nmcdiff} by dividing both sides by
$\bar d-\delta \bar d +\bar u \equiv \bar u (r_2+1)$, approximating
$\bar d-\delta \bar d +\bar u$ on the left hand side of Eq. \ref{nmcdiff}
by a parametrization of  $\bar d + \bar u$ and solving for $r_2$.
The structure functions and the parton distribution
are integrated for each data point
over  the same $Q^2$ regions as in the analysis of the charge ratio.
The result is shown in Fig. \ref{fig5} as solid triangles.
If we subtract the contribution of CSV from the ratio $r_2$ we obtain
the result shown as open triangles in Fig. \ref{fig5}.


\section{Conclusions}

In conclusion, we have examined in detail the discrepancy at
small $x$ between the CCFR neutrino and NMC muon structure
functions.  The only way we can make these data compatible is by assuming
charge symmetry violation in the sea quark distributions.   The
CSV amplitudes necessary to obtain agreement with experiment
are extremely large -- at least two orders
of magnitude greater than theoretical predictions of charge
symmetry violation. 
If CSV effects of this magnitude are really present, then one must
include charge symmetry violating quark distributions in phenomenological
models from the outset, and re-analyze
the extraction of parton distributions.

\section*{Acknowledgements}
This work is supported in part by the Australian Research Council,
and by the National Science Foundation under research contract
NSF-PHY9722706.

\end{document}